\documentclass[preprints,article,accept,moreauthors]{Definitions/mdpi} 

\firstpage{1} 
\pubvolume{1}
\issuenum{1}
\articlenumber{0}
\pubyear{2024}
\copyrightyear{2024}
\datereceived{ } 
\daterevised{ } 
\dateaccepted{ } 
\datepublished{ } 
\hreflink{https://doi.org/} 

\Title{Constraints on graviton mass from Schwarzschild precession in the orbits of S-stars around the Galactic Center}

\TitleCitation{Constraints on graviton mass from Schwarzschild precession in the orbits of S-stars around the Galactic Center}

\Author{Predrag Jovanovi\'{c}$^{1,*}$\orcidA{}, Vesna Borka Jovanovi\'{c}$^{2}$\orcidB{}, Du\v{s}ko Borka$^{2}$\orcidC{} and Alexander F. Zakharov$^{3}$\orcidD{}}

\AuthorNames{Predrag Jovanovi\'{c}, Vesna Borka Jovanovi\'{c}, Du\v{s}ko Borka and Alexander F. Zakharov}

\AuthorCitation{Jovanovi\'{c}, P.; Borka Jovanovi\'{c}, V.; Borka, D.; Zakharov, A. F. }

\address{%
$^{1}$ \quad Astronomical Observatory, Volgina 7, P.O. Box 74, 11060 Belgrade, Serbia \\
$^{2}$ \quad Department of Theoretical Physics and Condensed Matter Physics (020), Vin\v{c}a Institute of Nuclear Sciences - National Institute of the Republic of Serbia, University of Belgrade, P.O. Box 522, 11001 Belgrade, Serbia \\
$^{3}$ \quad Bogoliubov Laboratory for Theoretical Physics, JINR, 141980 Dubna, Russia \\
}

\corres{Correspondence: pjovanovic@aob.rs}

\abstract{In this paper we use a modification of the Newtonian gravitational potential with a non-linear Yukawa-like correction, as it was proposed by C. Will earlier to  obtain new bounds on graviton mass from the observed orbits of S-stars around the Galactic Center (GC). This phenomenological potential differs from the gravitational potential obtained in the weak field limit of Yukawa gravity, which we used in our previous studies. We also assumed that the orbital precession of S-stars is close to the prediction of General Relativity (GR) for Schwarzschild precession, but with a possible small discrepancy from it. This assumption is motivated by the fact that the GRAVITY Collaboration in 2020 and in 2022 detected Schwarzschild precession in the S2 star orbit around the Supermassive Black Hole (SMBH) at the GC. Using this approach, we were able to constrain parameter $\lambda$ of the potential and, assuming that it represents the graviton Compton wavelength, we also found the corresponding upper bound of graviton mass. The obtained results were then compared with our previous estimates, as well as with the estimates of other authors.}


\begin{document}

\section{Introduction}

Here we use a phenomenological modification of the Newtonian gravitational potential with
a non-linear Yukawa-like correction, as provided in \cite{will98,will18}, with the aim of obtaining new constraints on
graviton mass from the observed orbits of S-stars around the Galactic Center.

The modified theories of gravity have been suggested as alternative approaches to Newtonian gravity in order to explain astrophysical observations on different astronomical and cosmological scales. There are a significant number of theories
of modified gravity: \cite{fisc99,cope04,clif06,capo11a,capo11b,noji11,capo12a,clif12,noji17,salu21}.

Graviton is the gauge boson of the gravitational interaction and in GR theory is considered as massless, moving along null geodesics at the speed of light $c$. On the other hand, according to a class of alternative theories, known as  theories of massive gravity, gravitational interaction is propagated by a massive field, in which case graviton has some small non zero mass \cite{Fierz_39,Boulware_72,Logunov_88,Chugreev_89,Gershtein_03,Gershtein_04,Gershtein_06,ruba08,babi10,deRham_11,rham14,rham17}. This approach was first introduced in 1939 by Fierz and Pauli \cite{Fierz_39}.

The LIGO and Virgo collaborations considered a theory of massive gravity to be an appropriate approach and presented their estimate of the mass of the graviton $m_g < 1.2\times 10^{-22}~$eV in the first publication about the gravitational wave detection from binary black holes \cite{LIGO_16}. Analyzing signals observed during the three observing runs collected in The third Gravitational-wave Transient Catalog (GWTC-3) the LIGO -- Virgo  -- KAGRA collaborations obtained a stringer constraint of $m_g < 1.27\times 10^{-23}~$eV \cite{LIGO_21}. Various experimental constraints on the graviton mass are given in \cite{zyla20}.

There is a wide range of massive gravity theories, which lead to various phenomenologies \cite{rham17}.
However, several of such models predict that gravitational potential in the Newtonian limit acquires a Yukawa
suppression \cite{rham17}, so that the Poisson equation for Newtonian gravity $\nabla ^{2}\Phi =4\pi G\rho$ is
modified by graviton mass $m_g$ and, as noted in \cite{pois14}, it then takes the following form:
\begin{equation}
\left(\nabla^2+\dfrac{1}{\lambda^2}\right)\Phi=4\pi\,G\rho,
\label{eq:poisson}
\end{equation}
in which
\begin{equation}
\lambda = \dfrac{h}{m_g\,c}
\label{eq:compton}
\end{equation}
is the Compton wavelength of the graviton. In such a case (see e.g. \cite{will98,will18,pois14}), the spherically symmetric
potential $\Phi$ of a body of mass $M$ is given by 
\begin{equation}
\Phi\left(r\right)=-\dfrac{GM}{r}\, e^{-\dfrac{r}{\lambda}}.
\label{eq:potential}
\end{equation}
Different Yukawa-like potentials are analysed in papers: \cite{sand84,talm88,whit01,amen04,reyn05,seal05,moff05,moff06,sere06,capo07b,iori07,iori08,capo09b,adel09,card11}
and recently in \cite{rham17,will18,miao19,capo21,mart21,moni22,beni22b,dong22,tan24}.
As noted in \cite{will98}, the exact form of the Yukawa-like potential $\Phi$ should be, in principle, derived in the
frame of a complete theory of massive gravity. Therefore, in our previous investigations we studied the case of massive gravity obtained from $f(R)$ theories (see e.g. \cite{bork16,zakh16a,jova21,jova23,jova24}) which resulted with the Yukawa-like potential $\Phi$ with two parameters: the range of Yukawa interaction $\Lambda$ add its strength $\delta$. In contrast, here we follow the approach from \cite{will98} and assume the above mentioned particular phenomenology
according to which the potential $\Phi$ takes the form of Eq. (\ref{eq:potential}), regardless of the theoretical model
that produces it. As a consequence, we also assume that the metric at leading order in the Newtonian regime is
(see e.g. \cite{bern19}):
\begin{equation}
ds^2 = \left( -1 + \dfrac{2GM}{c^2 r}~e^{-\dfrac{r}{\lambda}} \right) c^2 dt^2 + \left( 1 + \dfrac{2GM}{c^2 r}~e^{-\dfrac{r}{\lambda}} \right) dl^2,
\quad dl^2 \equiv dx^2+dy^2+dz^2.
\label{eq:metric}
\end{equation}

Here we study the trajectories of S-stars orbiting around the central SMBH of our Galaxy, in the frame of Yukawa gravity using the modified PPN formalism \cite{clif08,alsi12,gain20a,gain20b}. Our present research is the continuation of our previous investigations of different Extended Gravity theories, where we used astrometric observations for S-star orbits \cite{bork12,bork13,zakh14,capo14,bork16,zakh16a,zakh18,dial19,bork19,jova21,bork21a,zakh22a,zakh22b,zakh23,bork22a,jova23,jova24}.

The compact radio source Sgr A$^\ast$ is very bright and located at the GC, while the so called S-stars are the bright stars which move around it. The orbits of these S-stars around Sgr A$^\ast$, which is recently confirmed to be a SMBH (as it was expected earlier \cite{lynd71,oort77,rees82,genz87,town90}), are monitored for about 30 years
{\cite{ghez00,scho02,ghez08,gill09a,gill09b,genz10,meye12,gill17,hees17,hees17b,chu18,abut18,abut19,do19,amor19,said19,hees20,abut20,genz22}. A number of analysis of S-star orbits was performed using available observational data by some theoretical groups (see e.g. \cite{doku15,mart18,dela18,kali20,lalr21,dadd21,lalr22,beni22a,beni23,bamb24}).

This paper is organized as follows: in Section 2 we presented the orbital precession in Yukawa-like gravitational potential. In Section 3 we introduced PPN equations of motion, together with the other important expressions that we used for analysis of the stellar orbits around Sgr A* in Yukawa gravity. We then performed an analysis for the potential from \cite{will18} that we have previously developed for some other modified potentials, and obtained the results for upper bound on graviton mass in the case of more different S-stars. These results are presented and discussed in section 4. Section 5 is devoted to the concluding remarks.

\section{Orbital precession in Yukawa-like gravitational potential}

In order to derive the expression for orbital precession in the gravitational potential (\ref{eq:potential}), we assume that
it does not differ significantly from the Newtonian potential $\Phi_N(r)=-\dfrac{GM}{r}$. Namely, it is well known that orbital
precession $\Delta\varphi$ per orbital period, induced by small perturbations to the Newtonian gravitational potential which
are described by the perturbing potential $V(r)=\Phi(r)-\Phi_N(r)$, could be evaluated as (see e.g. \cite{zakh18} and
references therein):
\begin{equation}
\Delta\varphi^{rad} = \dfrac{-2L}{GM e^2}\int\limits_{-1}^1
{\dfrac{z\cdot dz}{\sqrt{1 - z^2}}\dfrac{dV\left( z \right)}{dz}},
\label{eq:precorb}
\end{equation}
\noindent where $r$ is related to $z$ via $r = \dfrac{L}{1 + ez}$ and $L = a\left( {1 - {e^2}} \right)$ is the semilatus
rectum of the orbital ellipse. Approximate formula for orbital precession $\Delta\varphi_{\scriptscriptstyle Y}$ can be
obtained by performing the power series expansion of the perturbing potential $V(r)$ and by inserting the first order term
of this expansion into Eq. (\ref{eq:precorb}), which results with:
\begin{equation}
\Delta\varphi_{\scriptscriptstyle Y}^{rad}\approx\pi\sqrt{1-e^2}\dfrac{a^2}{\lambda^2}, \qquad a\ll\lambda.
\label{eq:precykw}
\end{equation}
Note that a similar expression for orbital precession was obtained in \cite{will18} and it was used for bounding the
Compton wavelength and mass of the graviton by the Solar System data.

\section{Stellar orbits in extended/modified PPN formalisms}

We simulated the orbits of S-stars around GC using the parameterized post-Newtonian (PPN) equations of motion (for more
details about PPN approach see e.g. \cite{will18b} and references therein). However, it is well known that Yukawa-like
potentials could not be entirely represented by the standard PPN formalism and thus require its extension/modification (see the related references in \cite{jova23}). This is also valid for the potential (\ref{eq:potential}) and its corresponding
metric (\ref{eq:metric}). Moreover, since Yukawa gravity is indistinguishable from GR up to the first
post-Newtonian correction \cite{clif08}, in addition to the standard PPN equations of motion $\vec{\ddot{r}}_{\scriptscriptstyle GR}$
in GR, PPN equations of motion $\vec{\ddot{r}}_{\scriptscriptstyle Y}$ in potential (\ref{eq:potential}) also include an
additional term $\vec{\ddot{r}}_{\scriptscriptstyle\lambda}$ with exponential correction due to the perturbing potential
$V(r)$. In this extended PPN formalism (denoted here as $\mathrm{PPN_Y}$), the equations of motion are:

\begin{equation}
\vec{\ddot{r}}_{\scriptscriptstyle Y} = \vec{\ddot{r}}_{\scriptscriptstyle GR} + \vec{\ddot{r}}_{\scriptscriptstyle\lambda},
\qquad \vec{\ddot{r}}_{\scriptscriptstyle GR} = \vec{\ddot{r}}_{\scriptscriptstyle N} + \vec{\ddot{r}}_{\scriptscriptstyle PPN},
\label{eq:eom}
\end{equation}

\noindent where the $\vec{\ddot{r}}_{\scriptscriptstyle N}$ is the Newtonian acceleration, $\vec{\ddot{r}}_{\scriptscriptstyle PPN}$
is the first post-Newtonian correction and $\vec{\ddot{r}}_{\scriptscriptstyle\lambda}$ is additional Yukawa correction given by
the following expressions, respectively:
\begin{equation}
\begin{array}{l}
\vec{\ddot{r}}_{\scriptscriptstyle N} = -GM \dfrac{\vec{r}}{r^3} \\
\\
\vec{\ddot{r}}_{\scriptscriptstyle PPN} = \dfrac{GM}{c^2r^3} \left[\left(4\dfrac{G M}{r}-\vec{\dot{r}}\cdot\vec{\dot{r}}\right) \vec{r} + 4\left(\vec{r}\cdot\vec{\dot{r}}\right)\vec{\dot{r}}\right] \\
\\
\vec{\ddot{r}}_{\scriptscriptstyle\lambda} = GM \left[ 1 - \left(1 + \dfrac{r}{\lambda} \right) e^{-\dfrac{r}{\lambda}} \right] \dfrac{\vec{r}}{r^3}.
\end{array}
\label{eq:ppn}
\end{equation}
The additional Yukawa correction $\vec{\ddot{r}}_{\scriptscriptstyle\lambda}$ becomes negligible when $\lambda\rightarrow\infty$,
and then $\vec{\ddot{r}}_{\scriptscriptstyle Y}\rightarrow\vec{\ddot{r}}_{\scriptscriptstyle GR}$, i.e. PPN equations
of motion $\vec{\ddot{r}}_{\scriptscriptstyle Y}$ in potential (\ref{eq:potential}) reduce to the standard PPN
equations of motion in GR. Therefore, the orbits of S-stars in Yukawa gravity and GR can be then simulated by numerical
integration of the corresponding expressions (\ref{eq:eom}).

The orbital precession in GR is given by the well known expression for the Schwarzschild precession \cite{will14}:
\begin{equation}
\Delta\varphi_{\scriptscriptstyle GR}^{rad}\approx\dfrac{6\pi G M}{c^2 a(1-e^2)},
\label{eq:precgr}
\end{equation}
where $a$ is semi-major axis and $e$ eccentricity of the orbit. Recently the GRAVITY Collaboration detected the orbital
precession of the S2 star around the SMBH at GC and showed that it was close to the above prediction of GR \cite{abut20}.
For that purpose, they introduced an ad hoc factor $f_{SP}$ in front of the first post-Newtonian correction of GR in order to parametrize the effect of the Schwarzschild metric. In this modified PPN formalism (denoted here as $\mathrm{PPN_{SP}}$),
the equations of motion are given by:
\begin{equation}
\vec{\ddot{r}}_{\scriptscriptstyle SP} = \vec{\ddot{r}}_{\scriptscriptstyle N} + f_{\scriptscriptstyle SP}\cdot\vec{\ddot{r}}_{\scriptscriptstyle PPN}.
\label{eq:eomfsp}
\end{equation}
The corresponding modified expression for the Schwarzschild precession is \cite{abut20}:
\begin{equation}
\Delta\varphi_{\scriptscriptstyle SP}^{rad}=f_{\scriptscriptstyle SP}\cdot\Delta\varphi_{\scriptscriptstyle GR}^{rad}.
\label{eq:precfsp}
\end{equation}
For $f_{SP}=1$ the expression (\ref{eq:eomfsp}) reduces to the standard PPN equations of motion $\vec{\ddot{r}}_{\scriptscriptstyle GR}$
in GR given in Eq. (\ref{eq:eom}), while the expression (\ref{eq:precfsp}) also reduces to the corresponding GR prediction
from Eq. (\ref{eq:precgr}). The parameter $f_{SP}$ shows to which extent some gravitational model is relativistic and it is defined as $f_{SP} = (2 + 2\gamma - \beta) / 3$ , where $\beta$ and $\gamma$ are the post-Newtonian parameters, and in the case of GR the both of them equals to 1 (and thus $f_{SP}=1$ in this case). 
For $f_{SP}=0$ the Newtonian case is recovered.
However, in the case of S2 star the value of $f_{SP} = 1.10 \pm 0.19$ was obtained by the GRAVITY Collaboration, indicating
the orbital precession of $f_{SP} \times 12.'1$ in its orbit around Sgr A* \cite{abut20}. Recently, this collaboration updated
the above first estimate to $f_{SP} = 0.85 \pm 0.16$, also obtained from the detected Schwarzschild precession of S2 star
orbit \cite{abut22}. Besides, they also presented the following estimate obtained from the fit with the four star (S2, S29,
S38, S55) data with the $1\sigma$ uncertainty: $f_{SP} = 0.997 \pm 0.144$ \cite{abut22}.

\section{Results: constraints on the Compton wavelength and mass of the graviton}

The constraints on parameter $\lambda$  under which the orbital precession in the gravitational potential (\ref{eq:potential})
deviates from the Schwarzschild precession in GR by a factor $f_{SP}$, can be obtained provided that the total orbital
precession in Yukawa gravity, given by the sum $\Delta\varphi_{\scriptscriptstyle GR}+\Delta\varphi_{\scriptscriptstyle Y}$,
is close to the observed precession $\Delta\varphi_{\scriptscriptstyle SP}$ obtained by the GRAVITY Collaboration:
\begin{equation}
\Delta\varphi_{\scriptscriptstyle Y} + \Delta\varphi_{\scriptscriptstyle GR} \approx \Delta\varphi_{\scriptscriptstyle SP}
\quad\Leftrightarrow\quad
\pi\sqrt{1-e^2}\dfrac{a^2}{\lambda^2} + \dfrac{6\pi GM}{c^2 a(1 - e^2)} \approx f_{SP}\dfrac{6\pi GM}{c^2 a(1 - e^2)}.
\label{eq:prectot}
\end{equation}
Taking into account the third Kepler law (since the orbits are almost Keplerian):
\begin{equation}
\dfrac{P^2}{a^3} \approx \dfrac{4\pi^2}{GM},
\label{eq:kepler}
\end{equation}
then from Eqs. (\ref{eq:prectot}) and (\ref{eq:kepler}) one can obtain the following relation between $\lambda$ and $f_{SP}$:
\begin{equation}
\lambda(P,e,f_{SP}) \approx \dfrac{cP}{2\pi}\dfrac{(1-e^2)^{\frac{3}{4}}}{\sqrt{6(f_{SP} - 1)}},\qquad f_{SP} > 1.
\label{eq:lambda}
\end{equation}
The above condition can be used for constraining the Compton wavelength $\lambda$ of the graviton by the observed values of
$f_{SP}$ only in the cases when $f_{SP}$ is larger than 1, since $\Delta\varphi_{\scriptscriptstyle Y}$ always gives a
positive contribution to the total precession in Eq. (\ref{eq:prectot}). The corresponding constraints on the graviton mass
$m_g$ can be then found by inserting the obtained value of $\lambda$ into Eq. (\ref{eq:compton}). The relative error of
the parameter $\lambda$ (and thus of the graviton mass $m_g$) in this case can be found by differentiating the
logarithm of the above expression (\ref{eq:lambda}):
\begin{equation}
\dfrac{|\Delta\lambda|}{\lambda}=\dfrac{|\Delta m_g|}{m_g} \leq \left(\dfrac{\left|\Delta P\right|}{P}+\dfrac{3 e\left|\Delta e\right|}{2 (1-e^2)}+\dfrac{\left|\Delta f_{SP}\right|}{2 (f_{SP}-1)}\right).
\label{eq:relerr}
\end{equation}
It can bee seen that the potential (\ref{eq:potential}) results with the same relative errors as the corresponding Yukawa potential derived in the frame of $f(R)$
theories of gravity (see e.g. \cite{jova24}).

\begin{table}[ht!]
\centering
\caption{The Compton wavelength of the graviton $\lambda$, its mass $m_g$, as well as their relative and absolute errors,
calculated for three different values of $f_{SP}$ in the case of S2 star.}
\label{tab1}
\setlength{\tabcolsep}{0.12cm}
\begin{tabular}{|c|c|rcl|rcl|r|}
\hline
$f_{SP}$&$\Delta f_{SP}$&\multicolumn{3}{c|}{$\lambda\pm\Delta\lambda$}&\multicolumn{3}{c|}{$m_g\pm\Delta m_g$}&\multicolumn{1}{c|}{R.E.} \\
& &\multicolumn{3}{c|}{(AU)}&\multicolumn{3}{c|}{$(10^{-24}\ \mathrm{eV})$}&\multicolumn{1}{c|}{(\%)} \\
\hline
\hline
1.100 & 0.190 & 66361.5 & $\pm$ & 63890.7 & 124.9 & $\pm$ & 120.2 &  96.3 \\
1.010 & 0.160 & 209853.4 & $\pm$ & 1681506.5 &  39.5 & $\pm$ & 316.5 & 801.3 \\
1.141 & 0.144 & 55886.4 & $\pm$ & 29251.3 & 148.3 & $\pm$ & 77.6 &  52.3 \\
\hline
\end{tabular}
\end{table}

We first used three estimates for $f_{SP}$ obtained by the GRAVITY Collaboration in order to find the corresponding constraints
on the Compton wavelength $\lambda$ of the graviton and its mass $m_g$ in the case of S2 star. These are the values of
$f_{SP}$ detected by GRAVITY collaboration in the case of S2 star \cite{abut20,abut22}, as well as from the combination
of a few stars: S2, S29, S38 and S55\cite{abut22}. For two estimates which are
lower than 1 we used the upper limits of their $1\sigma$ intervals, i.e. the values $f_{SP}+\Delta f_{SP}$. The obtained
constraints are given in Table \ref{tab1}, from which it can be seen that the most reliable result was obtained in the case
of $f_{SP}$ with the lowest uncertainty, resulting from the fit with the four star data. In that case the relative error
for $\lambda$ and $m_g$ was the lowest. In contrast, the worst constraint with unrealistically high relative error was obtained
in the second case with the lowest value of $f_{SP}=1.01$, due to the fact that it is too close to the corresponding
prediction of GR. By comparing the results obtained in the first case with our previous corresponding estimates from Table I in
\cite{jova24}, obtained for Yukawa-like potential derived from $f(R)$ theories of gravity, it can be seen that the upper
bound on graviton mass $m_g$ and its absolute error $\Delta m_g$ were improved for $\sim 30\%$ in the case of the
phenomenological potential (\ref{eq:potential}), although the relative error remained the same.

\begin{figure}[ht!]
\centering
\includegraphics[width=\linewidth]{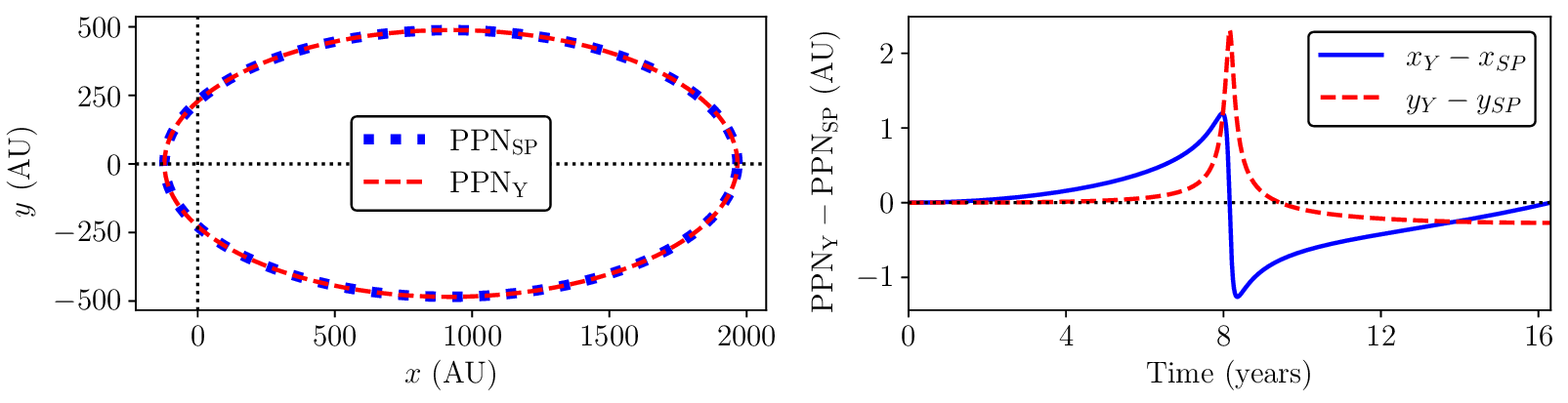}\\
\includegraphics[width=\linewidth]{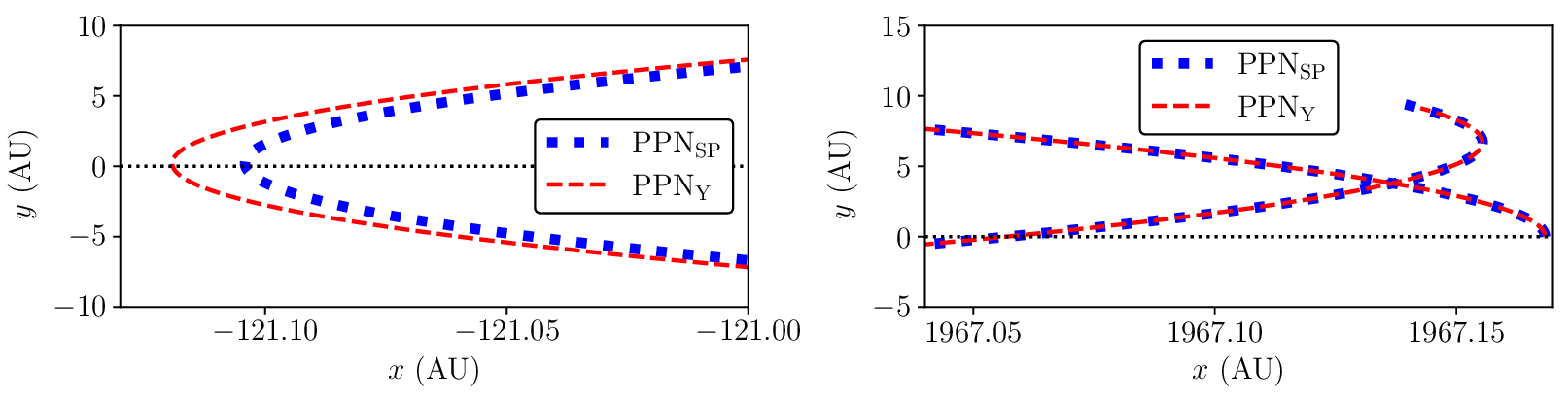}\\
\includegraphics[width=\linewidth]{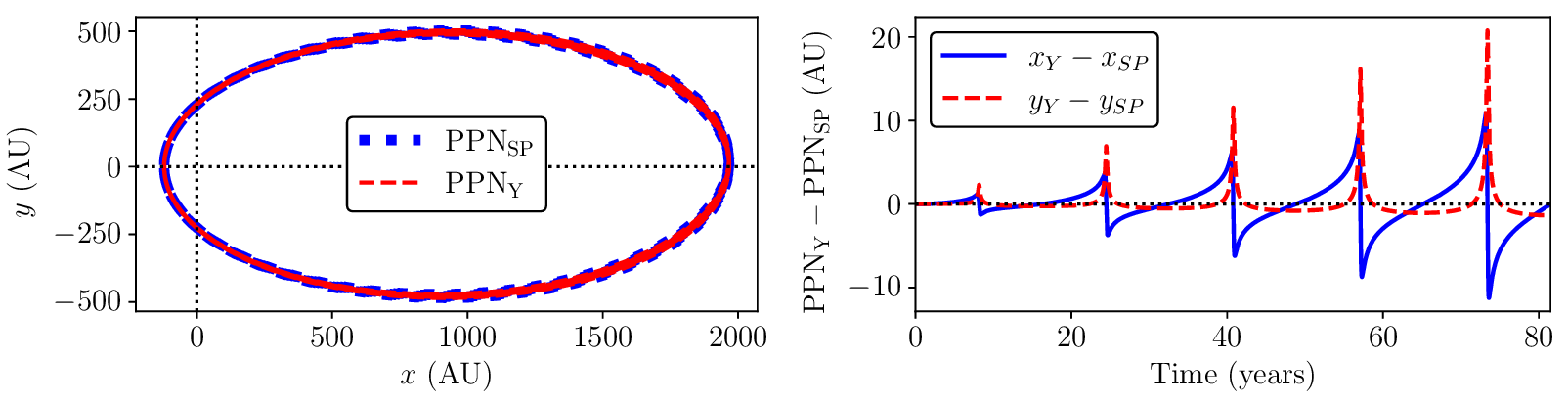}
\caption{\textit{Top left:} Comparison between the simulated orbits of S2 star during one orbital period, obtained by numerical integration of the equations of motion given by Eq. (\ref{eq:eomfsp}) in $\mathrm{PPN_{SP}}$ formalism for $f_{SP}=1.10$ (blue dotted line) and those given by Eq. (\ref{eq:eom}) in $\mathrm{PPN_Y}$ formalism for $\lambda=66361.5$ AU (red dashed line), which corresponds to $f_{SP}=1.10$ according to Eq. (\ref{eq:lambda}). \textit{Top right:} The differences between the corresponding $x$ and $y$ coordinates in $\mathrm{PPN_Y}$ and $\mathrm{PPN_{SP}}$ formalisms as the functions of time. \textit{Middle:} Comparison between the simulated orbits of S2 star, near pericenter (left) and near apocenter (right) in two PPN formalisms. \textit{Bottom:} The same as in top panel, but for five orbital periods.}
\label{fig01}
\end{figure}

In the first case of $f_{SP} = 1.10 \pm 0.19$ we also graphically compared the simulated orbits of S2 star, obtained by
numerical integration of equations of motion in $\mathrm{PPN_{SP}}$ formalism given by Eq. (\ref{eq:eomfsp}) with those
in $\mathrm{PPN_Y}$ formalism given by Eq. (\ref{eq:eom}) for $\lambda=66361.5$ AU which corresponds to this $f_{SP}$
according to Eq. (\ref{eq:lambda}). The comparisons during one and five orbital periods are presented in the top left and bottom left panels of Fig. \ref{fig01}, respectively, while in the corresponding right panels we presented the differences between the corresponding $x$ and $y$ coordinates in these two PPN formalisms as the functions of time. In order to see more clear the difference between the simulated orbits of S2 star in two PPN formalisms, we show in middle panel of Fig. \ref{fig01} part of its orbit near the pericenter (left panel) and near the apocenter (right panel).

As it can be seen from Fig. \ref{fig01},
the differences between the simulated orbits of S2 star in these two PPN formalisms are very small, and the maximum
discrepancy between the corresponding coordinates during the first orbital period is only $\sim 2$ AU at the pericenter.
This discrepancy slowly increases with time during the successive orbital periods due to neglecting of the higher order
terms in power series expansion of the perturbing potential $V(r)$ in Eq. (\ref{eq:precorb}). One should also note that the two PPN formalisms give close, but not exactly the same epochs of pericenter passage, as it can be seen from top right panel of Fig. \ref{fig01}.

This sufficiently small difference between the simulated orbits in the two studied PPN formalisms confirms that relation (\ref{eq:lambda}) could be used for obtaining the constraints on the Compton wavelength $\lambda$ of the graviton and its mass $m_g$ from the latest estimates for $f_{SP}$ obtained by the GRAVITY Collaboration.

\begin{table}[ht!]
\centering
\caption{The Compton wavelength of the graviton $\lambda$, its mass $m_g$, as well as their relative and absolute errors,
calculated for $f_{SP}=1.10\pm0.19$ and $f_{SP}=1.19\pm0.19$ in the case of all S-stars from Table 3 in \cite{gill17}
except of S111.}
\label{tab2}
\setlength{\tabcolsep}{0.09cm}
\begin{tabular}{|l|rcl|rcl|r|rcl|rcl|r|}
\hline
&\multicolumn{7}{c|}{$f_{SP}=1.1\pm0.19$}&\multicolumn{7}{c|}{$f_{SP}=1.19\pm0.19$}\\
\cline{2-15}
\multicolumn{1}{|l|}{Star}&\multicolumn{3}{c|}{$\lambda\pm\Delta\lambda$}&\multicolumn{3}{c|}{$m_g\pm\Delta m_g$}&\multicolumn{1}{c|}{R.E.}&\multicolumn{3}{c|}{$\lambda\pm\Delta\lambda$}&\multicolumn{3}{c|}{$m_g\pm\Delta m_g$}&\multicolumn{1}{c|}{R.E.} \\
&\multicolumn{3}{c|}{(AU)}&\multicolumn{3}{c|}{$(10^{-24}\ \mathrm{eV})$}&\multicolumn{1}{c|}{(\%)}&\multicolumn{3}{c|}{(AU)}&\multicolumn{3}{c|}{$(10^{-24}\ \mathrm{eV})$}&\multicolumn{1}{c|}{(\%)} \\
\hline
\hline
S1   &   1.6e+06 & $\pm$ & 1.6e+06 &   5.1 & $\pm$ & 5.1 & 100.7 &   1.2e+06 & $\pm$ & 6.6e+05 &   7.0 & $\pm$ & 3.9 &  55.7 \\
S2   &   6.6e+04 & $\pm$ & 6.4e+04 & 124.9 & $\pm$ & 120.2 &  96.3 &   4.8e+04 & $\pm$ & 2.5e+04 & 172.1 & $\pm$ & 88.3 &  51.3 \\
S4   &   8.8e+05 & $\pm$ & 8.5e+05 &   9.4 & $\pm$ & 9.1 &  96.7 &   6.4e+05 & $\pm$ & 3.3e+05 &  13.0 & $\pm$ & 6.7 &  51.7 \\
S6   &   1.0e+06 & $\pm$ & 9.5e+05 &   8.3 & $\pm$ & 7.9 &  95.2 &   7.2e+05 & $\pm$ & 3.6e+05 &  11.5 & $\pm$ & 5.8 &  50.2 \\
S8   &   5.5e+05 & $\pm$ & 5.4e+05 &  15.0 & $\pm$ & 14.7 &  98.0 &   4.0e+05 & $\pm$ & 2.1e+05 &  20.6 & $\pm$ & 10.9 &  53.0 \\
S9   &   4.5e+05 & $\pm$ & 4.4e+05 &  18.6 & $\pm$ & 18.6 &  99.7 &   3.2e+05 & $\pm$ & 1.8e+05 &  25.7 & $\pm$ & 14.0 &  54.7 \\
S12  &   2.4e+05 & $\pm$ & 2.3e+05 &  34.9 & $\pm$ & 33.6 &  96.4 &   1.7e+05 & $\pm$ & 8.9e+04 &  48.1 & $\pm$ & 24.7 &  51.4 \\
S13  &   5.5e+05 & $\pm$ & 5.2e+05 &  15.1 & $\pm$ & 14.5 &  95.5 &   4.0e+05 & $\pm$ & 2.0e+05 &  20.9 & $\pm$ & 10.5 &  50.5 \\
S14  &   7.3e+04 & $\pm$ & 7.8e+04 & 114.1 & $\pm$ & 122.4 & 107.3 &   5.3e+04 & $\pm$ & 3.3e+04 & 157.2 & $\pm$ & 98.0 &  62.3 \\
S17  &   8.7e+05 & $\pm$ & 8.5e+05 &   9.5 & $\pm$ & 9.2 &  97.1 &   6.3e+05 & $\pm$ & 3.3e+05 &  13.1 & $\pm$ & 6.8 &  52.1 \\
S18  &   4.5e+05 & $\pm$ & 4.3e+05 &  18.4 & $\pm$ & 17.8 &  96.5 &   3.3e+05 & $\pm$ & 1.7e+05 &  25.4 & $\pm$ & 13.1 &  51.5 \\
S19  &   9.4e+05 & $\pm$ & 1.1e+06 &   8.8 & $\pm$ & 10.2 & 116.4 &   6.8e+05 & $\pm$ & 4.9e+05 &  12.1 & $\pm$ & 8.7 &  71.4 \\
S21  &   2.5e+05 & $\pm$ & 2.5e+05 &  33.3 & $\pm$ & 33.2 &  99.6 &   1.8e+05 & $\pm$ & 9.9e+04 &  45.9 & $\pm$ & 25.1 &  54.6 \\
S22  &   5.9e+06 & $\pm$ & 6.7e+06 &   1.4 & $\pm$ & 1.6 & 114.1 &   4.3e+06 & $\pm$ & 3.0e+06 &   1.9 & $\pm$ & 1.3 &  69.1 \\
S23  &   4.5e+05 & $\pm$ & 5.2e+05 &  18.5 & $\pm$ & 21.4 & 115.6 &   3.2e+05 & $\pm$ & 2.3e+05 &  25.5 & $\pm$ & 18.0 &  70.6 \\
S24  &   1.3e+06 & $\pm$ & 1.3e+06 &   6.6 & $\pm$ & 6.8 & 103.2 &   9.2e+05 & $\pm$ & 5.3e+05 &   9.1 & $\pm$ & 5.3 &  58.2 \\
S29  &   7.4e+05 & $\pm$ & 8.1e+05 &  11.1 & $\pm$ & 12.2 & 109.1 &   5.4e+05 & $\pm$ & 3.5e+05 &  15.4 & $\pm$ & 9.8 &  64.1 \\
S31  &   1.1e+06 & $\pm$ & 1.0e+06 &   7.8 & $\pm$ & 7.5 &  96.4 &   7.8e+05 & $\pm$ & 4.0e+05 &  10.7 & $\pm$ & 5.5 &  51.4 \\
S33  &   1.8e+06 & $\pm$ & 1.9e+06 &   4.7 & $\pm$ & 5.0 & 107.0 &   1.3e+06 & $\pm$ & 7.9e+05 &   6.5 & $\pm$ & 4.0 &  62.0 \\
S38  &   1.1e+05 & $\pm$ & 1.0e+05 &  76.9 & $\pm$ & 73.3 &  95.4 &   7.8e+04 & $\pm$ & 3.9e+04 & 106.0 & $\pm$ & 53.4 &  50.4 \\
S39  &   2.5e+05 & $\pm$ & 2.5e+05 &  33.2 & $\pm$ & 32.8 &  98.8 &   1.8e+05 & $\pm$ & 9.7e+04 &  45.8 & $\pm$ & 24.6 &  53.8 \\
S42  &   3.2e+06 & $\pm$ & 4.0e+06 &   2.6 & $\pm$ & 3.1 & 122.7 &   2.4e+06 & $\pm$ & 1.8e+06 &   3.5 & $\pm$ & 2.7 &  77.7 \\
S54  &   1.9e+06 & $\pm$ & 3.5e+06 &   4.4 & $\pm$ & 8.4 & 188.3 &   1.4e+06 & $\pm$ & 1.9e+06 &   6.1 & $\pm$ & 8.8 & 143.3 \\
S55  &   9.6e+04 & $\pm$ & 9.3e+04 &  86.5 & $\pm$ & 84.5 &  97.6 &   6.9e+04 & $\pm$ & 3.7e+04 & 119.3 & $\pm$ & 62.7 &  52.6 \\
S60  &   6.6e+05 & $\pm$ & 6.4e+05 &  12.6 & $\pm$ & 12.3 &  97.7 &   4.8e+05 & $\pm$ & 2.5e+05 &  17.4 & $\pm$ & 9.2 &  52.7 \\
S66  &   8.5e+06 & $\pm$ & 8.6e+06 &   1.0 & $\pm$ & 1.0 & 101.4 &   6.2e+06 & $\pm$ & 3.5e+06 &   1.3 & $\pm$ & 0.8 &  56.4 \\
S67  &   5.2e+06 & $\pm$ & 5.2e+06 &   1.6 & $\pm$ & 1.6 & 100.1 &   3.8e+06 & $\pm$ & 2.1e+06 &   2.2 & $\pm$ & 1.2 &  55.1 \\
S71  &   1.3e+06 & $\pm$ & 1.4e+06 &   6.4 & $\pm$ & 6.8 & 107.3 &   9.4e+05 & $\pm$ & 5.9e+05 &   8.8 & $\pm$ & 5.5 &  62.3 \\
S83  &   7.6e+06 & $\pm$ & 8.4e+06 &   1.1 & $\pm$ & 1.2 & 110.3 &   5.5e+06 & $\pm$ & 3.6e+06 &   1.5 & $\pm$ & 1.0 &  65.3 \\
S85  &   2.3e+07 & $\pm$ & 4.9e+07 &   0.4 & $\pm$ & 0.8 & 211.0 &   1.7e+07 & $\pm$ & 2.8e+07 &   0.5 & $\pm$ & 0.8 & 166.0 \\
S87  &   2.0e+07 & $\pm$ & 2.1e+07 &   0.4 & $\pm$ & 0.4 & 102.4 &   1.5e+07 & $\pm$ & 8.5e+06 &   0.6 & $\pm$ & 0.3 &  57.4 \\
S89  &   3.6e+06 & $\pm$ & 3.8e+06 &   2.3 & $\pm$ & 2.5 & 107.8 &   2.6e+06 & $\pm$ & 1.6e+06 &   3.2 & $\pm$ & 2.0 &  62.8 \\
S91  &   1.2e+07 & $\pm$ & 1.2e+07 &   0.7 & $\pm$ & 0.7 & 101.9 &   8.4e+06 & $\pm$ & 4.8e+06 &   1.0 & $\pm$ & 0.6 &  56.9 \\
S96  &   8.4e+06 & $\pm$ & 8.4e+06 &   1.0 & $\pm$ & 1.0 & 100.0 &   6.1e+06 & $\pm$ & 3.3e+06 &   1.4 & $\pm$ & 0.7 &  55.0 \\
S97  &   1.5e+07 & $\pm$ & 1.9e+07 &   0.6 & $\pm$ & 0.7 & 125.9 &   1.1e+07 & $\pm$ & 8.8e+06 &   0.8 & $\pm$ & 0.6 &  80.9 \\
S145 &   4.5e+06 & $\pm$ & 6.1e+06 &   1.9 & $\pm$ & 2.5 & 136.7 &   3.2e+06 & $\pm$ & 3.0e+06 &   2.6 & $\pm$ & 2.4 &  91.7 \\
S175 &   8.2e+04 & $\pm$ & 9.0e+04 & 101.4 & $\pm$ & 111.8 & 110.3 &   5.9e+04 & $\pm$ & 3.9e+04 & 139.7 & $\pm$ & 91.2 &  65.3 \\
R34  &   7.6e+06 & $\pm$ & 9.2e+06 &   1.1 & $\pm$ & 1.3 & 120.5 &   5.5e+06 & $\pm$ & 4.2e+06 &   1.5 & $\pm$ & 1.1 &  75.5 \\
R44  &   3.3e+07 & $\pm$ & 5.2e+07 &   0.2 & $\pm$ & 0.4 & 156.2 &   2.4e+07 & $\pm$ & 2.7e+07 &   0.3 & $\pm$ & 0.4 & 111.2 \\
\hline
\end{tabular}
\end{table}

\begin{table}[ht!]
\centering
\caption{The same as in Table \ref{tab2} but for $f_{SP}=1.16\pm0.16$ and $f_{SP}=1.144\pm0.144$.}
\label{tab3}
\setlength{\tabcolsep}{0.09cm}
\begin{tabular}{|l|rcl|rcl|r|rcl|rcl|r|}
\hline
&\multicolumn{7}{c|}{$f_{SP}=1.16\pm0.16$}&\multicolumn{7}{c|}{$f_{SP}=1.144\pm0.144$}\\
\cline{2-15}
\multicolumn{1}{|l|}{Star}&\multicolumn{3}{c|}{$\lambda\pm\Delta\lambda$}&\multicolumn{3}{c|}{$m_g\pm\Delta m_g$}&\multicolumn{1}{c|}{R.E.}&\multicolumn{3}{c|}{$\lambda\pm\Delta\lambda$}&\multicolumn{3}{c|}{$m_g\pm\Delta m_g$}&\multicolumn{1}{c|}{R.E.} \\
&\multicolumn{3}{c|}{(AU)}&\multicolumn{3}{c|}{$(10^{-24}\ \mathrm{eV})$}&\multicolumn{1}{c|}{(\%)}&\multicolumn{3}{c|}{(AU)}&\multicolumn{3}{c|}{$(10^{-24}\ \mathrm{eV})$}&\multicolumn{1}{c|}{(\%)} \\
\hline
\hline
S1   &   1.3e+06 & $\pm$ & 7.2e+05 &   6.4 & $\pm$ & 3.6 &  55.7 &   1.4e+06 & $\pm$ & 7.6e+05 &   6.1 & $\pm$ & 3.4 &  55.7 \\
S2   &   5.2e+04 & $\pm$ & 2.7e+04 & 158.0 & $\pm$ & 81.0 &  51.3 &   5.5e+04 & $\pm$ & 2.8e+04 & 149.9 & $\pm$ & 76.8 &  51.3 \\
S4   &   7.0e+05 & $\pm$ & 3.6e+05 &  11.9 & $\pm$ & 6.1 &  51.7 &   7.4e+05 & $\pm$ & 3.8e+05 &  11.3 & $\pm$ & 5.8 &  51.7 \\
S6   &   7.9e+05 & $\pm$ & 4.0e+05 &  10.5 & $\pm$ & 5.3 &  50.2 &   8.3e+05 & $\pm$ & 4.2e+05 &  10.0 & $\pm$ & 5.0 &  50.2 \\
S8   &   4.4e+05 & $\pm$ & 2.3e+05 &  18.9 & $\pm$ & 10.0 &  53.0 &   4.6e+05 & $\pm$ & 2.4e+05 &  17.9 & $\pm$ & 9.5 &  53.0 \\
S9   &   3.5e+05 & $\pm$ & 1.9e+05 &  23.6 & $\pm$ & 12.9 &  54.7 &   3.7e+05 & $\pm$ & 2.0e+05 &  22.3 & $\pm$ & 12.2 &  54.7 \\
S12  &   1.9e+05 & $\pm$ & 9.7e+04 &  44.1 & $\pm$ & 22.7 &  51.4 &   2.0e+05 & $\pm$ & 1.0e+05 &  41.8 & $\pm$ & 21.5 &  51.4 \\
S13  &   4.3e+05 & $\pm$ & 2.2e+05 &  19.2 & $\pm$ & 9.7 &  50.5 &   4.6e+05 & $\pm$ & 2.3e+05 &  18.2 & $\pm$ & 9.2 &  50.5 \\
S14  &   5.7e+04 & $\pm$ & 3.6e+04 & 144.3 & $\pm$ & 89.9 &  62.3 &   6.1e+04 & $\pm$ & 3.8e+04 & 136.9 & $\pm$ & 85.3 &  62.3 \\
S17  &   6.9e+05 & $\pm$ & 3.6e+05 &  12.0 & $\pm$ & 6.3 &  52.1 &   7.3e+05 & $\pm$ & 3.8e+05 &  11.4 & $\pm$ & 5.9 &  52.1 \\
S18  &   3.6e+05 & $\pm$ & 1.8e+05 &  23.3 & $\pm$ & 12.0 &  51.5 &   3.8e+05 & $\pm$ & 1.9e+05 &  22.1 & $\pm$ & 11.4 &  51.5 \\
S19  &   7.4e+05 & $\pm$ & 5.3e+05 &  11.1 & $\pm$ & 8.0 &  71.4 &   7.8e+05 & $\pm$ & 5.6e+05 &  10.6 & $\pm$ & 7.5 &  71.4 \\
S21  &   2.0e+05 & $\pm$ & 1.1e+05 &  42.2 & $\pm$ & 23.0 &  54.6 &   2.1e+05 & $\pm$ & 1.1e+05 &  40.0 & $\pm$ & 21.8 &  54.6 \\
S22  &   4.7e+06 & $\pm$ & 3.2e+06 &   1.8 & $\pm$ & 1.2 &  69.1 &   4.9e+06 & $\pm$ & 3.4e+06 &   1.7 & $\pm$ & 1.2 &  69.1 \\
S23  &   3.5e+05 & $\pm$ & 2.5e+05 &  23.4 & $\pm$ & 16.5 &  70.6 &   3.7e+05 & $\pm$ & 2.6e+05 &  22.2 & $\pm$ & 15.7 &  70.6 \\
S24  &   1.0e+06 & $\pm$ & 5.8e+05 &   8.3 & $\pm$ & 4.8 &  58.2 &   1.1e+06 & $\pm$ & 6.1e+05 &   7.9 & $\pm$ & 4.6 &  58.2 \\
S29  &   5.9e+05 & $\pm$ & 3.8e+05 &  14.1 & $\pm$ & 9.0 &  64.1 &   6.2e+05 & $\pm$ & 4.0e+05 &  13.4 & $\pm$ & 8.6 &  64.1 \\
S31  &   8.5e+05 & $\pm$ & 4.3e+05 &   9.8 & $\pm$ & 5.0 &  51.4 &   8.9e+05 & $\pm$ & 4.6e+05 &   9.3 & $\pm$ & 4.8 &  51.4 \\
S33  &   1.4e+06 & $\pm$ & 8.6e+05 &   6.0 & $\pm$ & 3.7 &  62.0 &   1.5e+06 & $\pm$ & 9.1e+05 &   5.6 & $\pm$ & 3.5 &  62.0 \\
S38  &   8.5e+04 & $\pm$ & 4.3e+04 &  97.3 & $\pm$ & 49.0 &  50.4 &   9.0e+04 & $\pm$ & 4.5e+04 &  92.3 & $\pm$ & 46.5 &  50.4 \\
S39  &   2.0e+05 & $\pm$ & 1.1e+05 &  42.0 & $\pm$ & 22.6 &  53.8 &   2.1e+05 & $\pm$ & 1.1e+05 &  39.8 & $\pm$ & 21.4 &  53.8 \\
S42  &   2.6e+06 & $\pm$ & 2.0e+06 &   3.2 & $\pm$ & 2.5 &  77.7 &   2.7e+06 & $\pm$ & 2.1e+06 &   3.1 & $\pm$ & 2.4 &  77.7 \\
S54  &   1.5e+06 & $\pm$ & 2.1e+06 &   5.6 & $\pm$ & 8.0 & 143.3 &   1.6e+06 & $\pm$ & 2.2e+06 &   5.3 & $\pm$ & 7.6 & 143.3 \\
S55  &   7.6e+04 & $\pm$ & 4.0e+04 & 109.5 & $\pm$ & 57.6 &  52.6 &   8.0e+04 & $\pm$ & 4.2e+04 & 103.8 & $\pm$ & 54.6 &  52.6 \\
S60  &   5.2e+05 & $\pm$ & 2.7e+05 &  16.0 & $\pm$ & 8.4 &  52.7 &   5.5e+05 & $\pm$ & 2.9e+05 &  15.2 & $\pm$ & 8.0 &  52.7 \\
S66  &   6.7e+06 & $\pm$ & 3.8e+06 &   1.2 & $\pm$ & 0.7 &  56.4 &   7.1e+06 & $\pm$ & 4.0e+06 &   1.2 & $\pm$ & 0.7 &  56.4 \\
S67  &   4.1e+06 & $\pm$ & 2.3e+06 &   2.0 & $\pm$ & 1.1 &  55.1 &   4.4e+06 & $\pm$ & 2.4e+06 &   1.9 & $\pm$ & 1.0 &  55.1 \\
S71  &   1.0e+06 & $\pm$ & 6.4e+05 &   8.1 & $\pm$ & 5.0 &  62.3 &   1.1e+06 & $\pm$ & 6.8e+05 &   7.6 & $\pm$ & 4.8 &  62.3 \\
S83  &   6.0e+06 & $\pm$ & 3.9e+06 &   1.4 & $\pm$ & 0.9 &  65.3 &   6.4e+06 & $\pm$ & 4.2e+06 &   1.3 & $\pm$ & 0.8 &  65.3 \\
S85  &   1.8e+07 & $\pm$ & 3.0e+07 &   0.5 & $\pm$ & 0.8 & 166.0 &   1.9e+07 & $\pm$ & 3.2e+07 &   0.4 & $\pm$ & 0.7 & 166.0 \\
S87  &   1.6e+07 & $\pm$ & 9.3e+06 &   0.5 & $\pm$ & 0.3 &  57.4 &   1.7e+07 & $\pm$ & 9.8e+06 &   0.5 & $\pm$ & 0.3 &  57.4 \\
S89  &   2.8e+06 & $\pm$ & 1.8e+06 &   3.0 & $\pm$ & 1.9 &  62.8 &   3.0e+06 & $\pm$ & 1.9e+06 &   2.8 & $\pm$ & 1.8 &  62.8 \\
S91  &   9.1e+06 & $\pm$ & 5.2e+06 &   0.9 & $\pm$ & 0.5 &  56.9 &   9.6e+06 & $\pm$ & 5.5e+06 &   0.9 & $\pm$ & 0.5 &  56.9 \\
S96  &   6.6e+06 & $\pm$ & 3.6e+06 &   1.2 & $\pm$ & 0.7 &  55.0 &   7.0e+06 & $\pm$ & 3.8e+06 &   1.2 & $\pm$ & 0.7 &  55.0 \\
S97  &   1.2e+07 & $\pm$ & 9.6e+06 &   0.7 & $\pm$ & 0.6 &  80.9 &   1.2e+07 & $\pm$ & 1.0e+07 &   0.7 & $\pm$ & 0.5 &  80.9 \\
S145 &   3.5e+06 & $\pm$ & 3.2e+06 &   2.4 & $\pm$ & 2.2 &  91.7 &   3.7e+06 & $\pm$ & 3.4e+06 &   2.2 & $\pm$ & 2.0 &  91.7 \\
S175 &   6.5e+04 & $\pm$ & 4.2e+04 & 128.2 & $\pm$ & 83.7 &  65.3 &   6.8e+04 & $\pm$ & 4.4e+04 & 121.6 & $\pm$ & 79.4 &  65.3 \\
R34  &   6.0e+06 & $\pm$ & 4.6e+06 &   1.4 & $\pm$ & 1.0 &  75.5 &   6.4e+06 & $\pm$ & 4.8e+06 &   1.3 & $\pm$ & 1.0 &  75.5 \\
R44  &   2.6e+07 & $\pm$ & 2.9e+07 &   0.3 & $\pm$ & 0.3 & 111.2 &   2.8e+07 & $\pm$ & 3.1e+07 &   0.3 & $\pm$ & 0.3 & 111.2 \\
\hline
\end{tabular}
\end{table}

Taking the above considerations into account, we then estimated the Compton wavelength $\lambda$ of the graviton, its mass $m_g$, and also their relative and absolute errors for all S-stars from Table 3 in \cite{gill17} except for S111 star.
For that purpose, and in order to avoid the cases when $f_{SP} < 1$, we adopted the same strategy as in \cite{jova24} and
assumed that the recent GRAVITY estimates for $f_{SP}$ are very close to the value in GR of $f_{SP}=1$, so that these
estimates represent a confirmation of GR within $1 \sigma$. Therefore, we constrained the graviton mass $m_g$ in the
particular cases when $f_{SP}=1+\Delta f_{SP}\pm\Delta f_{SP}$, i.e. for $f_{SP} = 1.19\pm0.19$, $f_{SP} = 1.16\pm0.16$
and $f_{SP} = 1.144\pm0.144$. As before, we used the expressions (\ref{eq:lambda}), (\ref{eq:relerr}) and
(\ref{eq:compton}) for this purpose, and the obtained results are presented in Tables \ref{tab2} and \ref{tab3}.
Besides, Table \ref{tab2} also contains the results for $f_{SP} = 1.10 \pm 0.19$ since, as shown in Table \ref{tab1},
this estimate is sufficiently larger than 1.

Using data from these tables, in Fig. \ref{fig02} we give the comparison of the estimated Compton wavelength $\lambda$
of the graviton, as well as for graviton mass upper bound, for four stars (S2, S29, S38, S55) which the GRAVITY
collaboration used for the newest estimation of $f_{SP}$. As it can be seen from Fig. \ref{fig02}, all constraints in
the case of S2, S38 and S55 star are approximately of the same order of magnitude ($\lambda\sim 10^5$ AU and
$m_g\sim 10^{-22}$ eV). The only exception is S29 star since it results with an order of magnitude larger values of
$\lambda$, and hence an order of magnitude smaller estimates for upper bound on graviton mass $m_g$. This is not
surprising because S29 star has much longer orbital period of $\sim 101$ yr with respect to the orbital periods of the
other three stars which are between $\sim 13-20$ yr (see Table 3 from \cite{gill17}).

\begin{figure}[ht!]
\centering
\includegraphics[width=0.48\textwidth]{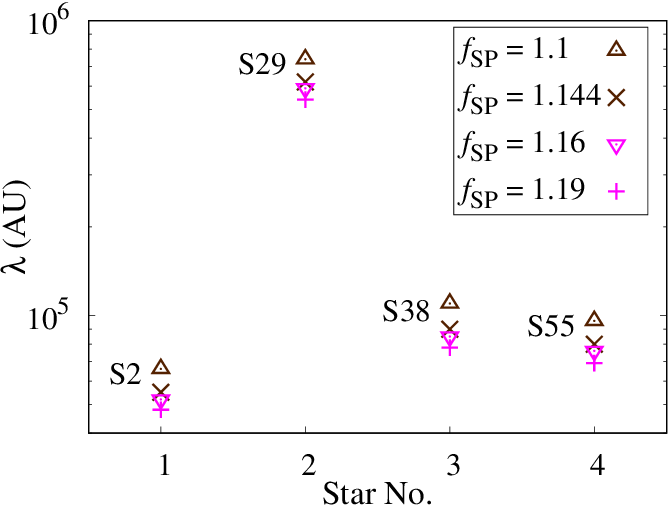}
\hfill
\includegraphics[width=0.48\textwidth]{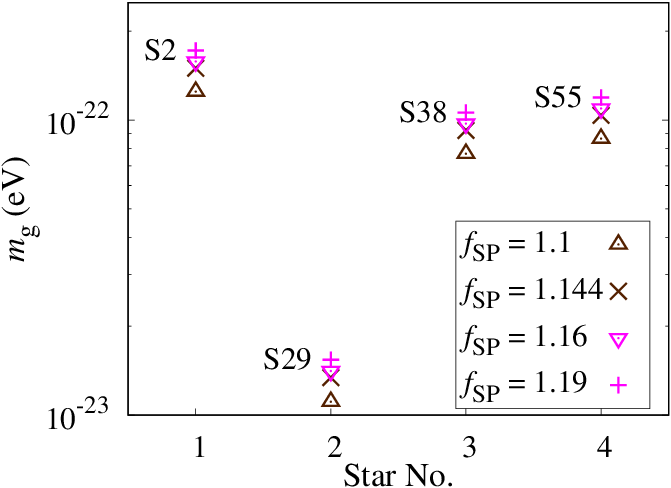}
\caption{\textit{Left:} Constraints on the Compton wavelength $\lambda$ of the graviton from the orbits of S2, S29, S38
and S55 star, in case of the following $f_{SP}$ estimates: 1.1, 1.144, 1.16 and 1.19. \textit{Right:} Upper bounds on
graviton mass $m_g$, for the same group of S-stars and the same $f_{SP}$.}
\label{fig02}
\end{figure}

If one compares these results with our previous corresponding estimates from Tables I and III in \cite{jova24}, obtained for Yukawa-like potential derived from $f(R)$ theories of gravity, it can be noticed that both upper bound on graviton mass
$m_g$ and its absolute error in the case of Yukawa-like potential (\ref{eq:potential}) were further improved for
$\sim 30\%$, in a similar way as for S2 star in the first case from Table \ref{tab1} (i.e. for $f_{SP} = 1.10 \pm 0.19$).

Although the current GRAVITY estimates of $f_{SP}=1.10\pm 0.19$ (from \cite{abut20}) and $f_{SP}=0.85\pm0.16$ and
$f_{SP}=0.997\pm0.144$ (from \cite{abut22}) can improve our previous constraints on the upper bound of graviton mass
for about $\sim 30\%$ (these results can be compared with our previous corresponding estimates from Tables I and III
in \cite{jova24} for the corresponding S-star), we have to stress that we take assumption that $f_{SP}$ has been
measured for all S-stars orbits already to a given precision. In reality, it is expected that the orbits of different S-stars should result with slightly different measured values and accuracies of $f_{SP}$. Because of that, our assumption that $f_{SP}$ is the same for all S-stars (see Tables \ref{tab2} and \ref{tab3}) probably does not hold.

\subsection{PPN fit of the observed orbit of S2 star}

In order to verify the results presented in Tables \ref{tab2} and \ref{tab3}, we also estimated the value of the Compton
wavelength $\lambda$ of the graviton by fitting the simulated orbits in the extended $\mathrm{PPN_Y}$ formalism into
the observed orbit of S2 star. For that purpose we used the publicly available astrometric observations of S2 star from
\cite{gill17}. Orbital fitting in the frame of extended $\mathrm{PPN_Y}$ formalism (\ref{eq:eom}) was performed by minimization
of the reduced $\chi^2$ statistics:
\begin{equation}
\chi_{\mathrm{red}}^2 = \dfrac{1}{2\left(N-\nu\right)}{\sum\limits_{i = 1}^N {\left[ {{{\left( {\dfrac{x_i^o - x_i^c}{\sigma_{xi}}}
\right)}^2} + {{\left( \dfrac{y_i^o - y_i^c}{\sigma_{yi}} \right)}^2}} \right]} },
\label{eq:chi2}
\end{equation}
where $(x_i^o, y_i^o)$ is the $i$-th observed position, $(x_i^c, y_i^c)$ is the corresponding calculated position, $N$
is the number of observations, $\nu$ is number of unknown parameters, $\sigma_{xi}$ and $\sigma_{yi}$ are the observed
astrometric uncertainties.

The values of the graviton Compton wavelength $\lambda$, SMBH mass $M$, distance $R$ to the GC and the osculating
orbital elements $a, e, i, \Omega, \omega, P, T$ which correspond to the minimum of $\chi_{\mathrm{red}}^2$ are found
using the differential evolution optimization method, implemented as Python Scipy function \href{https://docs.scipy.org/doc/scipy/reference/generated/scipy.optimize.differential_evolution.html}{\nolinkurl{scipy.optimize.differential\_evolution}}.
This is a population-based metaheuristic search technique of finding the global minimum of a multivariate function which
is especially suitable in evolutionary computations, since it is stochastic in nature, does not use gradient descent to
find the minimum, can search large areas of candidate space and seeks to iteratively enhances a candidate solution
concerning a specified quality metric. In order to improve the minimization slightly, the final result of the
differential evolution optimization is further polished at the end using Python Scipy function
\href{https://docs.scipy.org/doc/scipy/reference/generated/scipy.optimize.minimize.html}{\nolinkurl{scipy.optimize.minimize}}.
This also results with an approximation for the inverse Hessian matrix of $\chi_{\mathrm{red}}^2$, which on the other
hand could be considered as a good estimation for the covariance matrix of the parameters. Therefore, the standard error
for each fitted parameter can be calculated by taking the square root of the respective diagonal element of this
covariance matrix.

A particular value of $\chi_{\mathrm{red}}^2$ which corresponds to some specific combination of the mentioned adjustable
parameters is calculated in the following way:
\begin{enumerate}
\item
First, a simulated orbit of S2 star in the extended $\mathrm{PPN_Y}$ formalism is calculated by numerical integration of
the equations of motion (\ref{eq:eom}), starting from initial conditions $(x_0, y_0, \dot{x}_0, \dot{y}_0)$, where the
first two components specify the initial position and the last two the initial velocity in the orbital plane. In our
simulations the initial conditions correspond to the time of apocenter passage $t_\mathrm{ap}$ preceding the first
astrometric observation at $t_0$: $t_\mathrm{ap}=T-(2k-1)\dfrac{P}{2}$, where $T$ is the time of pericenter passage,
$P$ is the orbital period and $k$ is the smallest positive integer (number of periods) for which $t_\mathrm{ap}\le{t_0}$.
Then, the initial conditions are: $x_0=-r_\mathrm{ap}$, $y_0=0$, $\dot{x}_0=0$ and $\dot{y}_0=-v_\mathrm{ap},$ where
$r_\mathrm{ap}=a(1+e)$ is the apocenter distance and $v_\mathrm{ap}=\dfrac{2\pi\,a}{P}\sqrt{\dfrac{1-e}{1+e}}$ is the
corresponding orbital velocity at the apocenter.
\item
The true orbit obtained in the first step, which depends only on $a, e, P, T$, was then projected to the observer's sky
plane using the remaining geometrical orbital elements $i, \Omega, \omega$, in order to obtain the corresponding
positions $(x^c,y^c)$ along the apparent orbit:
\begin{equation}
x^c=Bx+Gy,\qquad y^c=Ax+Fy,
\label{eq:project}
\end{equation}
where $A,B,F,G$ are the Thiele-Innes elements:
\begin{equation}
\begin{array}{l}
A=\cos\Omega\cos\omega-\sin\Omega\sin\omega\cos i,\\
B=\sin\Omega\cos\omega+\cos\Omega\sin\omega\cos i,\\
F=-\cos\Omega\sin\omega-\sin\Omega\cos\omega\cos i,\\
G=-\sin\Omega\sin\omega+\cos\Omega\cos\omega\cos i.\\
\end{array}
\label{eq:abfg}
\end{equation}
In addition, the radial velocities $V_\mathrm{rad}$ are obtained from the corresponding true positions $(x,y)$ and
orbital velocities $(\dot{x},\dot{y})$ as (see e.g. \cite{bork13} and references therein):
\begin{equation}
V_\mathrm{rad} =  \dfrac{\sin i}{\sqrt{x^2 + y^2}} \left[ \sin(\theta + \omega) \cdot(x \dot{x} + y \dot{y}) +
\cos (\theta + \omega) \cdot (x \dot{y} - y \dot{x}) \right],
\label{eq:vrad}
\end{equation}
where $\theta = \arctan \dfrac{y}{x}$.
\item
Finally, $\chi_{\mathrm{red}}^2$ is obtained according to Eq. (\ref{eq:chi2}), taking into account only those apparent
positions $(x^c,y^c)$ which are calculated at the same epochs as the astrometric observations $(x^o,y^o)$.
\end{enumerate}

The obtained results of the orbital fitting in the case of S2 star are presented in Fig. \ref{fig03}, and the corresponding
best-fit values of the parameters are given in Table \ref{tab4}. As it can be seen from Fig. \ref{fig03}, and since
the best fit resulted with $\chi_{\mathrm{red}}^2=1.108$ which is only slightly larger than 1, the best-fit orbit of S2
star in the extended $\mathrm{PPN_Y}$ formalism is in a very good agreement with the observations.

\begin{table}[ht!]
\centering
\caption[]{Best-fit values of the graviton Compton wavelength $\lambda$, SMBH mass $M$, distance $R$ to the GC and
the osculating orbital elements $a, e, i, \Omega, \omega, P, T$ of the S2 star orbit.}
\label{tab4}
\begin{tabular}{lllll}
\hline
\noalign{\smallskip}
Parameter& Value & Fit error & Unit \\
\noalign{\smallskip}
\hline
\noalign{\smallskip}
$\lambda$& 82175.7 & 9828.05 & AU \\
$M$      & 4.15 & 0.27 & $10^6\,M_\odot$ \\
$R$      & 8.33 & 0.198 & kpc \\
$a$      & 0.1229 & 0.00430 & arcsec \\
$e$      & 0.8797 & 0.01597 & \\
$i$      & 134.89 & 1.984 & $^\circ$ \\
$\Omega$ & 224.57 & 5.208 & $^\circ$ \\
$\omega$ & 62.78 & 4.562 & $^\circ$ \\
$P$      & 15.98 & 0.362 & yr \\
$T$      & 2018.12219 & 0.696709 & yr \\
\noalign{\smallskip}
\hline
\noalign{\smallskip} \noalign{\smallskip}
\end{tabular}
\end{table}

\begin{figure}[ht!]
\centering
\includegraphics[width=\textwidth]{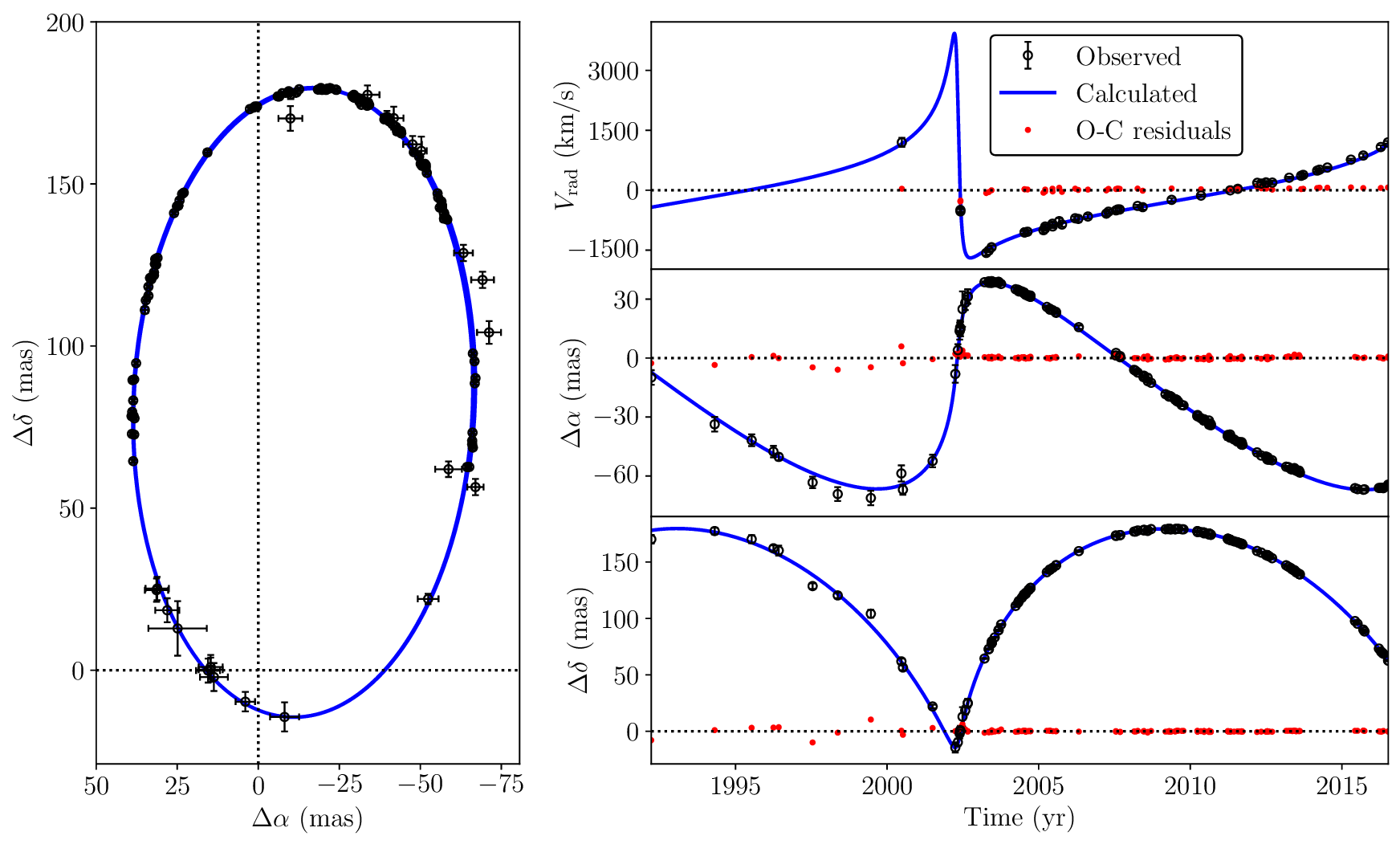}
\caption{\textit{Left:} Comparison between the best-fit orbit of the S2 star (blue solid line), simulated in the extended
$\mathrm{PPN_Y}$ formalism, and the corresponding astrometric observations from \cite{gill17} (black circles with error bars).
\textit{Right:} The same for the radial velocity of the S2 star (top), as well as for its $\alpha$ (middle) and $\delta$
(bottom) offset relative to the position of Sgr A* at the coordinate origin. Red dots in the right panels denote the
corresponding O-C residuals.}
\label{fig03}
\end{figure}

Concerning the graviton Compton wavelength $\lambda$ obtained from S2 star orbit, it can be seen that its best-fit value
of $\lambda\approx 8.2\times 10^4$ AU from Table \ref{tab4} is a bit larger, but still within the error intervals of the
corresponding values from Tables \ref{tab2} and \ref{tab3}, obtained according to Eq. (\ref{eq:lambda}) from the
detected values of $f_{SP}$. Therefore, the results of direct orbital fitting are in agreement with our constraints on
the graviton Compton wavelength $\lambda$ and mass $m_g$ presented in Tables \ref{tab2} and \ref{tab3}, which were
obtained from the values of $f_{SP}$ estimated by the GRAVITY Collaboration.

\section{Conclusions}

Here we used the phenomenological Yukawa-like gravitational potential from \cite{will98,will18} in order to obtain the
constraints on the graviton mass $m_g$ from the detected Schwarzschild precession in the observed stellar orbits around
the SMBH at GC. For that purpose we used two modified/extended PPN formalisms in order to derive the relation between the
Compton wavelength $\lambda$ of the graviton and parameter $f_{SP}$ which parametrizes the effect of the Schwarzschild
metric, and which was obtained by the GRAVITY Collaboration from the observed stellar orbits at GC. The results from this
study can be summarized as follows:
\begin{enumerate}
\item
We found the condition for parameter $\lambda$ of the phenomenological Yukawa-like gravitational potential (\ref{eq:potential})
under which the orbital precession in this potential deviates from the Schwarzschild precession in GR by a factor $f_{SP}$;
\item
The relation (\ref{eq:lambda}) derived from the phenomenological potential (\ref{eq:potential}) in the frame of the two
modified/extended PPN formalisms could be used for obtaining the reliable constraints on the graviton mass $m_g$ from the
latest estimates for $f_{SP}$ by the GRAVITY Collaboration in the cases when $f_{SP} > 1$;
\item
Both studied PPN formalisms result with close and very similar simulated orbits of S-stars, which practically overlap during
the first orbital period and then begin to slowly diverge over time due to some assumed theoretical approximations;
\item
In most cases, the constraints on the upper bound on graviton mass $m_g$ and its absolute error $\Delta m_g$, obtained
using the phenomenological potential (\ref{eq:potential}) were improved for $\sim 30\%$ in respect to our previous
corresponding estimates from \cite{jova24}, obtained using a slightly different Yukawa-like potential derived in the frame
of $f(R)$ theories of gravity, although the relative errors in both cases remained the same;
\item
These results were also confirmed in the case of S2 star by fitting of its observed orbit in the frame of the extended
$\mathrm{PPN_Y}$ formalism, which resulted with the best-fit value for the graviton Compton wavelength $\lambda$ within
the error intervals of its corresponding estimates obtained according to Eq. (\ref{eq:lambda}) from the detected values of $f_{SP}$;
\item
The least reliable constraints with unrealistically high uncertainties were only obtained from the estimates for $f_{SP}$
which were very close to its value predicted by GR, being just slightly larger than 1;
\item
If one compares the results from Table \ref{tab2} with those from Table \ref{tab3}, it can be seen that the upper bounds on graviton mass $m_g$ are very similar. In the case of S2 star $m_g < (1.5\pm 0.8)\times 10^{-22}~$eV and relative error is around 50\%. We can conclude that the more precise future observations are required in order to further improve the upper graviton mass bounds.
\end{enumerate}

\authorcontributions{All coauthors participated in writing, calculation and discussion of obtained results.}

\acknowledgments{The Authors acknowledge the support of Ministry of Science, Technological Development and Innovations of the Republic of Serbia through the Project contracts No. 451-03-66/2024-03/200002 and 451-03-66/2024-03/200017.}

\conflictsofinterest{The authors declare no conflict of interest.} 

\abbreviations{Abbreviations}{The following abbreviations are used in this manuscript:\\
\noindent
\begin{tabular}{@{}ll}
GC & Galactic Center \\
GR & General Relativity \\
SMBH & Supermassive black hole \\
PPN formalism & parameterized post-Newtonian formalism \\
\end{tabular}}

\reftitle{References}

\end{document}